# Design of an SI Engine Cold Start Controller based on Dynamic Coupling Analysis

Mohammad Reza Amini and Mahdi Shahbakhti

Department of Mechanical Engineering-Engineering Mechanics,
Michigan Technological University, Houghton, Michigan, USA

**Abstract:** In this paper, the dynamic couplings among different inputs and outputs of a highly nonlinear spark ignition (SI) engine control problem during the cold start phase are evaluated by using relative gain array (RGA) technique. First, based on the experimental data, a multi-input multi-output model is developed to represent the engine dynamics. Second, the coupling among different inputs and outputs of the model is evaluated by using RGA technique in both open-loop and closed-loop structures. The results show that although there is an internal coupling within the engine dynamics in the open-loop framework, the closed-loop engine controller can be designed using a decentralized structure without significantly affecting the system performance. In the next step, based on a nonlinear physics-based model of the engine, a set of single-input single-output (SISO) adaptive second order discrete sliding mode controllers (DSMC) are designed to drive the states of the engine model to their pre-defined desired trajectories and minimize the tailpipe HC emission, under modelling and implementation (data sampling and quantization) uncertainties. The real-time test results on an actual engine control unit (ECU) show that the proposed SISO adaptive second order DSMC provides accurate and fast tracking performance for the highly nonlinear and internally coupled engine dynamics, and can meet the HC emission limit by controlling the engine-out emissions and exhaust catalytic converter efficiency.

## 1-Introduction

Cold start phase in spark ignition (SI) engines has significant effects on the produced carbon monoxide (CO) and unburnt hydrocarbon (HC) emissions, fuel economy, and vehicle drivability. The 5-10 minute time period between the cold start and the moment the engine's coolant temperature reaches 80-100°C, is defined as the cold phase [1]. The generated HC emission during the cold start phase is responsible for over 80% of the total emissions in standard driving cycles [2,3]. When the ambient temperature drops to $-7^oC$, HC emission levels will increase by $2.5-4$ times, in comparison to the case with $20^oC$ ambient temperature [4]. The cold start phase is an unavoidable part of standard driving cycles. Thus, reducing the tailpipe HC emission during the cold start phase is an important challenge with increasing importance as the industry moves towards green vehicles.

There are two main reasons for high HC emission production rate during the cold start phase (i) the instability of the combustion inside the cylinders due to poor fuel-air mixing, cold cylinders walls, and poor evaporation of cold fuel, and (ii) low conversion efficiency of catalytic converters at low temperatures. Unlike the active emission reduction methods [5-7], in which extra equipment is added either to the engine or to the exhaust aftertreatment system to meet the desired requirements, in the passive HC emission reduction methods, the main component to minimize the tailpipe emission is the exhaust catalytic converter. At low temperatures, due to the low conversion efficiency of catalytic converters, the exhaust aftertreatment system cannot mitigate unburnt HC until catalyst "light-off" (i.e., operating temperature of about $300^oC$) is reached.

The optimum performance of the exhaust oxidation catalyst is achieved if the engine can follow the desired operation trajectories [1]. One way to increase the exhaust gas temperature for heating-up the catalyst and reducing the catalytic converter light-off time is retarding the spark timing; however, this results in higher engine-out HC emission. The other parallel approach to heat-up the catalyst is increasing the exhaust gas flow rate by increasing the idle engine speed. Increasing the idle engine speed with fixed injected fuel amount leads to lean air-to-fuel ratio (*AFR*), which is not desirable during the beginning of cold start phase, because of limited fuel evaporation and mixing that can lead to unstable combustion. Thus, more fuel needs to be injected to keep the *AFR* rich enough during the cold start phase. Overall, adjusting engine cold start operation is a complex nonlinear control problem that requires trade-offs among engine speed, exhaust gas temperature, and *AFR* controls.

High level of nonlinearity and complexity in the engine cold start transient dynamics call for nonlinear model-based control techniques. The first requirement for developing a model-based engine cold start controller is having an accurate and computationally efficient physics-based model which can estimate the transient engine dynamics. The second requirement is that the designed controller should be easily verifiable and can be easily implemented on a real engine. Additionally, the desired controller should be robust and has an adaptive structure against external and internal sources of uncertainties. The major sources of uncertainties are (i) implementation imprecisions introduced by the analog-to-digital converter (ADC) unit via data sampling and quantization, and (ii) uncertainties within the plant model due to aging, unseen dynamics, and model parameter variations for the transient conditions that the model has not been calibrated for. Thereby, in this paper, a nonlinear control-oriented model of an SI engine coupled with a three-way catalytic converter is used to design a model-based nonlinear controller to regulate the tailpipe HC emission. A second order adaptive discrete sliding mode controller (DSMC) framework is used in this study to design the engine cold start controller, since the second order adaptive DSMC is a robust and computationally efficient controller design technique that can treat systems with nonlinear dynamics with a great deal of uncertainty [8]. Eventually, in order to verify the designed controller performance in real-time, the cold start DSMC is tested in a processor-in-the-loop (PIL) setup by using an actual engine control unit (ECU), under modeling and implementation imprecisions.

The internally coupled and complex dynamics of the engine cold start require a controller framework with multiple input and outputs. The second order DSMC can be designed in both decentralized (single-input single-output (SISO)) and centralized (multi-input multi-output (MIMO)) structures. This means that the engine cold start controller can be designed with a single MIMO or a set of SISO DSMCs. Design of a decentralized controller has several benefits [9,10], including (i) improved productivity through module reuse, (ii) easy integration of new features, (iii) enhanced maintainability, and (iv) module sharing across powertrain platforms. On the other side, the calibration time and efforts for a set of decentralized controllers can be cumbersome, compared to a single MIMO controller. The tailpipe HC emission during the cold start phase is a function of several inputs. Thus, for the closed-loop controller design, the level of interaction among different control loops between several input/outputs (I/Os) pairs should be analyzed to understand the closed-loop dynamics coupling. The purpose is to determine proper pairings between engine cold start model I/Os to match the plant inputs and outputs that have the largest effect on each other within the DSMC structure [11]. Relative gain array (RGA) is a well-recognized dynamic coupling analysis method which provides a systematic approach to determine the coupling between controller I/Os [11]. The RGA method is used in this paper to conduct closed-loop dynamics coupling analysis and determine the proper structure for the second order DSMC, either SISO (decentralized) or MIMO (centralized).

The contribution of this paper is twofold. First, the RGA technique is used to perform a closed-loop dynamics coupling analysis to understand the structure of the closed-loop system. The RGA analysis requires a linear model which is derived based on the available experimental data for a 2.4-Liter SI engine during the cold start phase. Second, according to the RGA analysis results, a SISO adaptive second order DSMC is designed and experimentally tested in real-time for tracking the desired engine trajectories, in order to minimize the cold start tailpipe HC emission.

## 2- Engine Cold Start Dynamic Coupling Analysis

This section determines the engine dynamic coupling and required controller structure (SISO or MIMO) for the second order DSMC by performing RGA analysis. This will be done by finding the proper pairings between different inputs and outputs of the plant/controller, and eventually match up those pairs of I/O which have the largest effect on each. The RGA analysis is done for a model of the engine with three inputs ($u_i$) and three outputs ($y_i$):

$$Y = \begin{bmatrix} y_1 \\ y_2 \\ y_3 \end{bmatrix} = \begin{bmatrix} AFR \\ \omega_e \\ T_{exh} \end{bmatrix} \tag{1}$$

$$U = \begin{bmatrix} u_1 \\ u_2 \\ u_3 \end{bmatrix} = \begin{bmatrix} \dot{m}_{fc} \\ \dot{m}_{ai} \\ \Delta \end{bmatrix} \tag{2}$$

where $\omega_e$ is the engine speed, $T_{exh}$ is the exhaust gas temperature, $\dot{m}_{fc}$ is the mass flow rate of injected fuel into the cylinders, $\dot{m}_{ai}$ is the air mass flow rate into the intake manifold, and $\Delta$ is the spark timing after top dead centre (ATDC). The RGA matrix ($\mathbf{\Lambda}$), for the system with Eq. (1) and (2), is defined as:

$$\mathbf{\Lambda} = \begin{bmatrix} \lambda_{11} & \lambda_{12} & \lambda_{13} \\ \lambda_{21} & \lambda_{22} & \lambda_{23} \\ \lambda_{32} & \lambda_{32} & \lambda_{33} \end{bmatrix} \tag{3}$$

where $\lambda_{ij}$ is the relative gain between $y_i$ and $u_j$, and is obtained as follows [11,12]:

$$\lambda_{ij} = \frac{p_{ij}}{q_{ij}} \tag{4}$$

in which, $p_{ij}$ is the open-loop gain between $y_i$ and $u_j$, and $q_{ij}$ is the closed-loop gain between $y_i$ and $u_j$. $p_{ij}$ gains show the level of interaction among different inputs and outputs of the engine model in an open-loop structure. The open-loop gain matrix ($\mathbf{P}$) is defined as [11]:

$$\mathbf{P} = \begin{bmatrix} p_{11} & p_{12} & p_{13} \\ p_{21} & p_{22} & p_{23} \\ p_{31} & p_{32} & p_{33} \end{bmatrix} = \begin{bmatrix} \dfrac{\delta AFR}{\delta \dot{m}_{fc}} & \dfrac{\delta AFR}{\delta \dot{m}_{ai}} & \dfrac{\delta AFR}{\delta \Delta} \\ \dfrac{\delta \omega_e}{\delta \dot{m}_{fc}} & \dfrac{\delta \omega_e}{\delta \dot{m}_{ai}} & \dfrac{\delta \omega_e}{\delta \Delta} \\ \dfrac{\delta T_{exh}}{\delta \dot{m}_{fc}} & \dfrac{\delta T_{exh}}{\delta \dot{m}_{ai}} & \dfrac{\delta T_{exh}}{\delta \Delta} \end{bmatrix} \tag{5}$$

where $p_{ij}$ can be determined by utilizing experimental tests. For instance, $p_{11}$ is found by applying a small change ($\delta \dot{m}_{fc}$) in $\dot{m}_{fc}$ and measure the impact of this change on the corresponding output ($\delta AFR$). This should be done when the engine runs at the steady state conditions and there are no changes in the other two inputs. Instead of doing the experimental tests which is time consuming and costly, here a more convenient approach is taken to find the open-loop gains. By looking into Eq. (5), it can be easily concluded that the open-loop matrix is equivalent to the linearized engine model with three inputs ($\dot{m}_{fc}, \dot{m}_{ai}, \Delta$) and three outputs ($AFR, \omega_e, T_{exh}$). Thus, a linear model of the engine is required to perform the RGA analysis. To this end, a linear model is identified based on the available engine cold start experimental data.

Figure 1 shows two sets of experimental data which are taken from the cold start operation of a 2.4-Liter, 4-cylinder, DOHC 16 valve Toyota 2AZ-FE engine with a close-coupled three-way catalytic converter. Details of the engine experimental setup are found in [13-15]. These two data sets are merged into the MATLAB® System Identification Toolbox to find first order continuous-time linear sub-models. The linear model has three inputs ($\dot{m}_{ai}, \dot{m}_{fc}, \Delta$) and three outputs ($\omega_e, AFR, T_{exh}$). The open-loop bode plots of the identified data-based first order sub-models are shown in Figure 2. As can be seen, all the outputs of the model are affected by all the inputs. This observed internal coupling from the identified model is consistent with the physics of the engine cold start dynamics. $AFR$ is mainly a function of the injected fuel ($\dot{m}_{fc}$) and air mass flow rate ($\dot{m}_{ai}$); however, the spark timing ($\Delta$) also affects the $AFR$. The link between $\Delta$ and $AFR$ can be traced in the coupling between engine torque, speed, and resulting air flow variations. In a similar manner to $AFR$ dynamics, engine speed dynamics are considerably affected by $\dot{m}_{fc}$ and $\dot{m}_{ai}$. Moreover, engine speed is affected by the spark timing through torque generation. Finally, as can be seen from the third row of the Figure 2, exhaust gas temperature dynamic is a strong function of the spark timing and fuel injection. The main control input for regulating the exhaust gas temperature is the spark timing; however, any changes in the fuel injection rate can impact the combustion phasing inside the cylinder and consequently alter the exhaust gas temperature. In addition, increasing the engine speed ($\omega_e$) leads to less heat loss inside the cylinder and consequently affects temperature of exhaust gases leaving the engine cylinder. This explains the link between $T_{exh}$ and $\omega_e$ ($\dot{m}_{ai}$). Table 1 and Figure 3 show the performance of the first order model in estimating the engine dynamics. Next, the identified linear sub-models are used in the RGA analysis for finding the open-loop gains in Eq. (5).

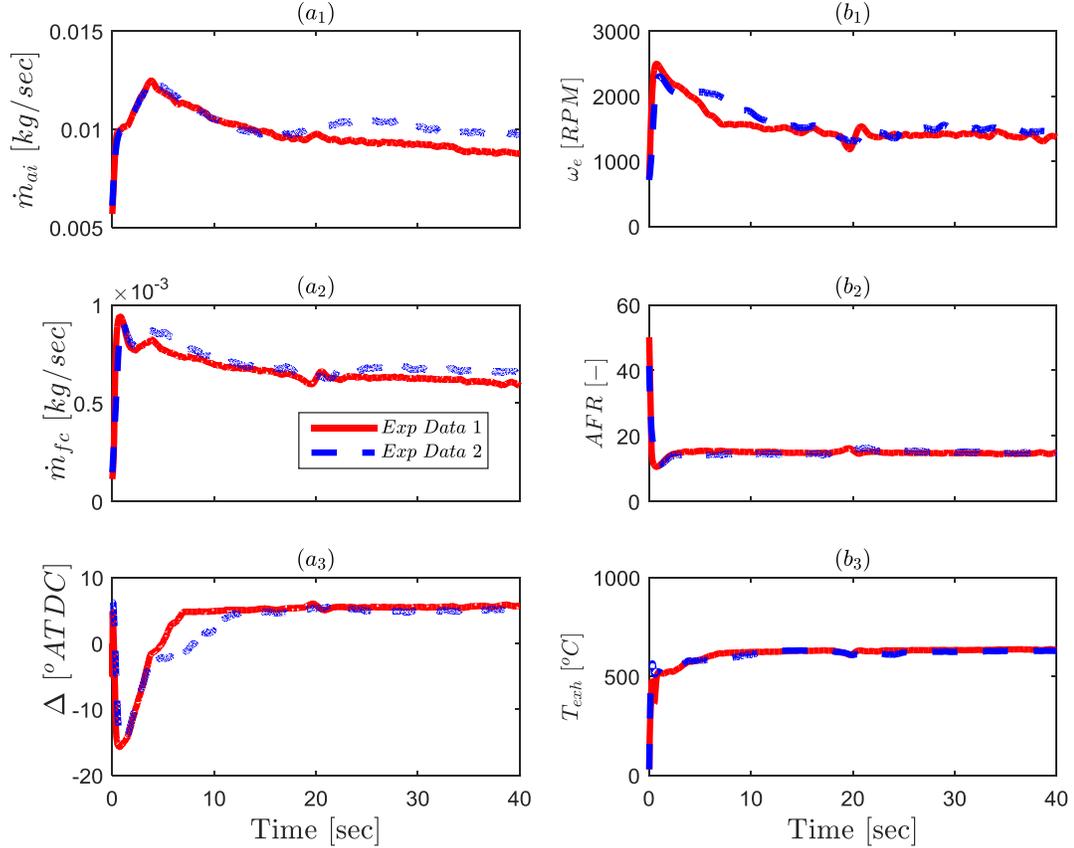

**Figure 1: Experimental data used for model identification, ($a_{1,2,3}$) inputs: $\dot{m}_{ai}, \dot{m}_{fc}, \Delta$,**
**($b_{1,2,3}$) outputs: $\omega_e, AFR, T_{exh}$.**

**Table 1: Mean ($\bar{e}$) and standard deviation ($\sigma_e$) errors in the first order estimated cold start engine model**
**compared to the experimental data.**

|  | $AFR$ $[-]$ | $\omega_e$ $[rad/sec]$ | $T_{exh}$ $[^oC]$ |
|---|---|---|---|
| $\bar{e}$ | 0.3 | 8.7 | 10.1 |
| $\sigma_e$ | 0.3 | 4.6 | 12.8 |

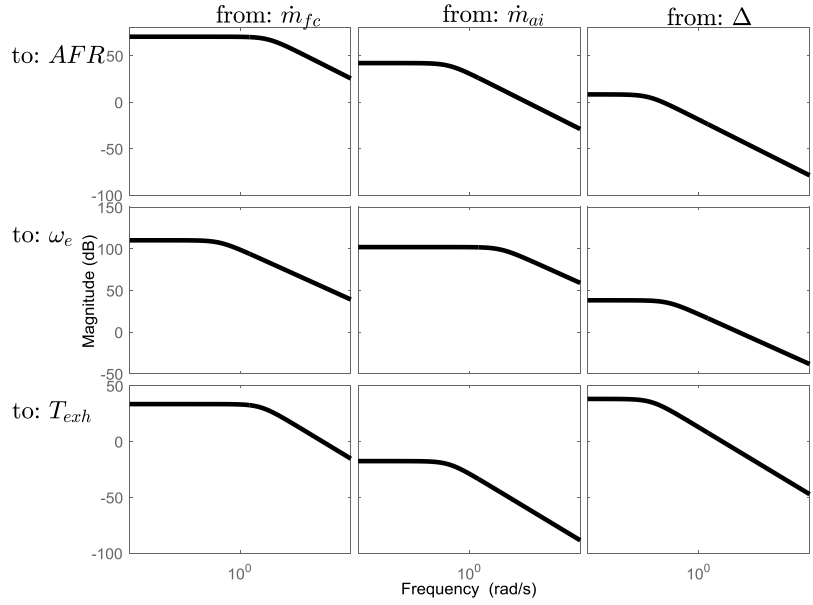

**Figure 2: Open-loop bode plot of the identified model based on the experimental data shown in Figure 1 with three inputs ($\dot{m}_{fc}, \dot{m}_{ai}, \Delta$) and three outputs ($AFR, \dot{m}_f, T_{exh}$).**

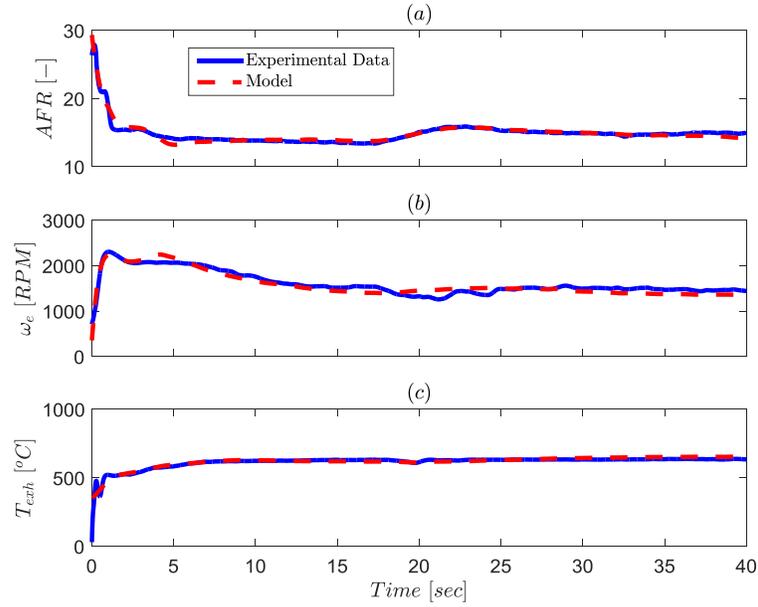

**Figure 3: Estimated first order model in comparison with the experimental data (a) $AFR$, (b) engine speed, and (c) exhaust gas temperature.**

In the frequency domain, the transfer function of the linear engine model has the same concept of the open-loop gains (Eq. (5)). Thus, in the frequency domain, **P** can be alternatively defined as:

$$\mathbf{P}(s) = \begin{bmatrix} G_{11}(s) & G_{12}(s) & G_{13}(s) \\ G_{21}(s) & G_{22}(s) & G_{23}(s) \\ G_{31}(s) & G_{32}(s) & G_{33}(s) \end{bmatrix}, \qquad G_{ij}(s) = \frac{1}{\tau_{ij}s + k_{ij}} \tag{6}$$

where, the bode plots of $\mathbf{P}(s)$ were already shown in Figure 2. According to [11,12], once the open-loop gains are found, the RGA matrix ($\mathbf{\Lambda}$), in the frequency domain, can be obtained by solving the following equation:

$$\mathbf{\Lambda}(s) = \mathbf{P}(s) \otimes \left(\mathbf{P}^{-1}(s)\right)^T \tag{7}$$

in which, $\left(\mathbf{P}^{-1}\right)^T$ is the transpose of the inverted $\mathbf{P}$. Additionally, "$\otimes$" denotes the element-by-element product. If it is assumed that $\left(\mathbf{P}^{-1}\right)^T$ has the following structure:

$$\left(\mathbf{P}^{-1}\right)^T = \begin{bmatrix} c_{11} & c_{12} & c_{13} \\ c_{21} & c_{22} & c_{23} \\ c_{31} & c_{32} & c_{33} \end{bmatrix} \tag{8}$$

then [11]:

$$c_{ij} = \frac{1}{q_{ij}} \tag{9}$$

where $q_{ij}$ is the closed-loop gain between $y_i$ and $u_j$. If the calculated RGA value is equal to zero ($\lambda_{ij} = 0$), it means that the open-loop gain between $y_i$ and $u_j$ is zero and there is no interaction between this pair of I/O. If $\lambda_{ij} = 1$ (equivalent to *20log(1)=0* in the bode plot), it can be concluded that $y_i$ and $u_j$ are strong function of each other, and in the closed-loop structure, the control loop between $y_i$ and $u_j$ is not affected by other loops. On the side, if $0 \ll \lambda_{ij} \ll 1$, then the closed-loop between $y_i$ and $u_j$ is affected by other inputs, and there is interaction among different control loops.

Figure 4 shows the results of the engine cold start dynamics coupling analysis by using the RGA technique based on the first order sub-models shown in Figure 2. The first important observation from Figure 4 is the dominant diagonal RGA values ($\lambda_{ii} = 1$, which represented by "0" in the bode plots). This means that despite the complex interaction among different inputs and outputs of the engine dynamics, *in a closed-loop structure* the diagonal pairing between inputs and outputs ($u_i \Leftrightarrow y_i$) are not affected by other loops ($u_j \Leftrightarrow y_i, i \neq j$). In other words, the closed-loop control system for the engine during the cold start phase strongly leans towards a decentralized (SISO) structure. It can be seen that for the *AFR* ($y_1$) controller, as expected, the main regulatory control input is the fuel injection rate ($u_1$). Although *AFR* is linked to air mass flow dynamics, the loop between *AFR* and $\dot{m}_{ai}$ has no considerable effect on the main *AFR* control loop with $\dot{m}_{fc}$. Moreover, the loop between *AFR* and $\Delta$ barely affects the *AFR* $\Leftrightarrow \dot{m}_{fc}$ control loop.

Similar to *AFR*, for the engine speed controller, the dominant loop is the $\omega_e \Leftrightarrow \dot{m}_{ai}$ (i.e., $y_2 \Leftrightarrow u_2$) loop, while with a much lower extend, the $\omega_e \Leftrightarrow \dot{m}_{fc}$ loop is affecting the main closed-loop. The reason for this impact from fuel injection on the engine speed is the link between *AFR* and intake air flow dynamics. Especially during the cold start phase, in which the *AFR* is first rich and then it becomes stoichiometric. Tracking the desired *AFR*, while idle engine speed usually varies during the cold start, explains how fuel injection is linked to the rotational dynamics. In a similar manner to $AFR \Leftrightarrow \dot{m}_{fc}$ loop, the $\omega_e \Leftrightarrow \Delta$ loop has no significant effect on the main engine speed closed-loop ($\omega_e \Leftrightarrow \dot{m}_{ai}$). Eventually, for the exhaust gas temperature controller, it can be seen that $T_{exh} \Leftrightarrow \Delta$ is the main closed-loop and the other two loops ($T_{exh} \Leftrightarrow \dot{m}_{fc}$ and $T_{exh} \Leftrightarrow \dot{m}_{ai}$) has no considerable impact on the $T_{exh} \Leftrightarrow \Delta$ loop. Overall, the results of dynamic coupling analysis by using the RGA method suggest to use a SISO structure for the engine cold start controller. Thus, in the next section, a set of three SISO DSMCs will be designed and tested for the engine cold start model.

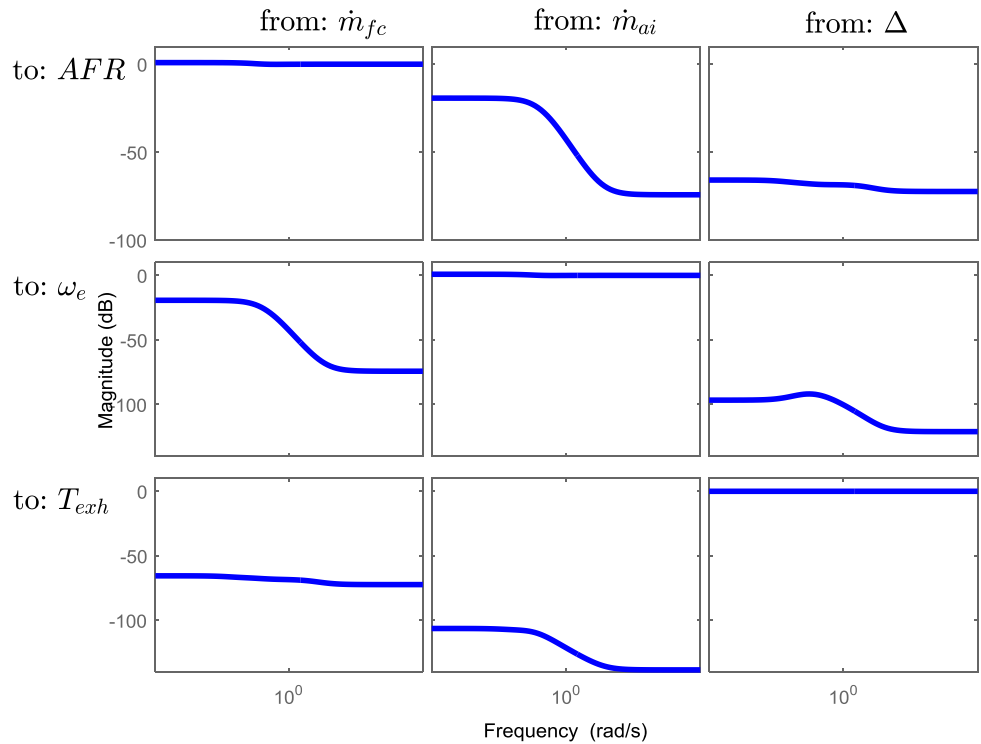

**Figure 4: Bode plots of RGA for the identified data-based engine model with three inputs ($\dot{m}_{ai}, \dot{m}_{fc}, \Delta$) and three outputs ($AFR, \omega_e, T_{exh}$).**

## 3-Design of SISO Model-Based Cold Start Controller

### 3-1-Engine Cold Start Emission Model

An accurate and computationally efficient physics-based model of the engine cold start is required that considers the complex and nonlinear dynamics of the SI engine during the cold start phase. To this end, a control-oriented cold start emission model from [13] is used in this section for designing a model-based controller. The model has two sub-models (i) SI engine model, and (ii) catalytic converter model. The first sub-model captures major dynamics affecting $AFR$, engine speed, and exhaust gas temperature. The engine model has five states: the air mass at intake manifold ($m_a$), the engine speed ($\omega_e$), the mass flow rate of injected fuel into cylinders ($\dot{m}_f$), the catalyst temperature ($T_{cat}$) and the exhaust gas temperature ($T_{exh}$). The model has three inputs: $\dot{m}_{ai}$, $\dot{m}_{fc}$, and $\Delta$. The engine sub-model is parameterized for the 2.4-Liter direct injection Toyota 2AZ-FE engine. Further modeling procedure and experimental validation of the model are detailed in [14] and [15]. The states and equations describing the model are as follows:

$$x = \begin{bmatrix} m_a \\ \omega_e \\ \dot{m}_f \\ T_{cat} \\ T_{exh} \end{bmatrix}, \quad \dot{x}(t) = \begin{bmatrix} \dot{m}_a \\ \dot{\omega}_e \\ \dot{m}_f \\ \dot{T}_{cat} \\ \dot{T}_{exh} \end{bmatrix} = \begin{bmatrix} \dot{m}_{ai} - \dot{m}_{ao} \\ \frac{1}{J}(\tau_{net} - \tau_L) \\ \frac{1}{\alpha_f}(\dot{m}_{fc} - \dot{m}_f) \\ \frac{1}{m_{cp}}(\dot{Q}_{gen} - \dot{Q}_{in} - \dot{Q}_{out}) \\ \frac{1}{\alpha_e}(ST \times AFI - T_{exh}) \end{bmatrix} \tag{10}$$

Details of the functions and the parameters in Eq. (10) are found in the Appendix. The second sub-model simulates the engine–out HC, the catalytic converter efficiency, and tailpipe HC emissions. The HC production rate from the engine ($\dot{HC}_{eng}$) is calculated along with the catalytic converter efficiency ($\eta_{cat}$) to give the HC emission rate out of the tailpipe ($\dot{HC}_{tp}$) [16]:

$$\dot{HC}_{eng}(t) = \frac{\dot{m}_f(r_c-1)}{r_c} e^{\left[-a\left(\frac{\theta_{EVO}-\theta_0}{\delta\theta}\right)^n\right]} \tag{11}$$

$$\eta_{cat}(t) = 0.98 \left(1 - exp\left[-5\left(\frac{\frac{AFR}{AFR_{st}} - 0.7}{0.3}\right)^{15}\right]\right) \cdot \left(1 - exp\left[-0.2\left(\frac{T_{cat} - 30}{150}\right)^5\right]\right) \tag{12}$$

$$\dot{HC}_{tp}(t) = \dot{HC}_{eng}(1 - \eta_{cat}) \tag{13}$$

where, $AFR_{st}$ indicates stoichiometric air fuel ratio, $r_c$ the compression ratio. $\delta\theta, \theta_0, \theta_{EVO}$ are parameters to calculate the fuel burn rate as detailed in the Appendix. Eq. (10) to (13) altogether shape the dynamical system equations in which the output of the model is the tailpipe HC flow rate. The difference between engine-out and tailpipe HC emissions depends on the catalytic converter performance in the exhaust system. The cold start model has been validated with experimental data from the same engine in the previous work [17]. The validation results in [17] show that at the end of the simulation period (200 seconds), there is less than 2% error between measured and simulated tailpipe HC emissions.

**3-2-Second Order Adaptive DSMC Design for Engine Emission Controls**

The main control objective is to reduce the cumulative HC tailpipe emission, which is a function of the engine transient performance and the catalytic converter efficiency. In this paper, desired $AFR_d$ and $\omega_d$ control trajectories are taken from the test results using the ECU of Toyota engine. The exhaust gas temperature of 650°C is chosen for the desired exhaust gas temperature using available experimental data for the engine [18]. It is assumed that if the engine can track the pre-defined desired trajectories accurately, then the catalyst light-off time will be shorten and the tailpipe emission will be mitigated effectively. This means that the engine cold start emission control problem is narrowed down to a tracking control problem. To this end, in this paper a second order discrete sliding mode control (DSMC) from [8] is chosen for cold start controller design. This selection is done due to (i) dealing with highly nonlinear dynamics of the engine cold start model, (ii) the uncertainties in the plant model, (iii) imprecisions at the controller inputs/outputs (I/O) due to data sampling and quantization, and (iv) high frequency oscillations due to chattering phenomena in conventional sliding mode controls. The structure of DSMC, as shown in our previous work [8], allows for deriving the adaptation laws to remove uncertainties within the engine model, and guarantees the closed-loop controller stability.

According to the dynamics coupling analysis results from Section 2, it was concluded that despite the internal coupling between different inputs and outputs of the engine, the engine controller, with $\dot{m}_{fc}, \dot{m}_{ai}, \Delta$ as the inputs and $AFR, \omega_e, T_{exh}$ as the outputs, strongly leans towards a decentralized (SISO) structure. Thus, in this section, a set of three SISO controllers are designed to track desired trajectories for $AFR$, $\omega_e$, and $T_{exh}$. Based on these three desired trajectories, three discrete first order sliding variables $(s)$, as the tracking errors at each time step $(k)$, are defined:

$$s_1(k) = AFR(k) - AFR_d(k) \tag{14}$$

$$s_2(k) = \omega_e(k) - \omega_{e,d}(k) \tag{15}$$

$$s_3(k) = T_{exh} - T_{exh,d}(k) \tag{16}$$

Next, the second order sliding variables ($\xi$) are defined with respect to the first order sliding variable as follows [19]:

$$\xi_i(k) = s_i(k) + \beta_i s_i(k), \;\; i = 1,2,3 \tag{17}$$

where, $\beta_i$ is the gain of the second order DSMC and it should be $1 > \beta_i > 0$ to guarantee the asymptotic stability of the closed-loop system [8]. The main difference between the first and second order DSMCs is that on the sliding manifold, in addition to the zero convergence of the sliding variable ($s$) which is satisfied by applying the first order DSMC, the second order DSMC drives the derivative (difference function) of $s$ to zero. The control input of the second order DSMC on the sliding manifold is obtained by solving the following equalities in discrete time:

$$\xi_i(k+1) = \xi_i(k) = 0, \;\; i = 1,2,3 \tag{18}$$

Eq. (18) is equivalent to the $s(t) = \dot{s}(t) = 0$ for a second order sliding mode controller in continuous-time domain [20]. In order to find the control inputs of the second order DSMC for the engine case study with respect to Eq. (18), first the continuous-time model of the engine in Eq. (10) should be discretized by using the Euler approximation method [21,22]. It should be noted from Eq. (10) that since there is no direct control input for regulating the engine speed, the desired air mass inside the intake manifold ($m_{a,d}$) is considered as the synthetic control input for the engine speed controller. The calculated $m_{a,d}$ is then used as the desired trajectory for the air mass flow controller which has $\dot{m}_{ai}$ as the control input. Thus, the first ($s_4$) and second ($\xi_4$) order sliding variables for the forth control surface are defined as follows:

$$s_4(k) = m_a(k) - m_{a,d}(k) \tag{19}$$

$$\xi_4(k) = s_4(k+1) + \beta_4 s_4(k) \tag{20}$$

Moreover, since $T_{cat}$ is not among the desired trajectories of the engine controller and its dynamics strongly depend on the dynamics of the catalytic converter, $T_{cat}$ is not considered in the controller formulation. Therefore, the discrete from of the continuous-time nonlinear model in Eq. (10), in the absence of $T_{cat}$ state, becomes [22]:

$$\begin{bmatrix} m_a(k+1) \\ \omega_e(k+1) \\ \dot{m}_f(k+1) \\ T_{exh}(k+1) \end{bmatrix} = \begin{bmatrix} m_a(k) \\ \omega_e(k) \\ \dot{m}_f(k) \\ T_{exh}(k) \end{bmatrix} + T \begin{bmatrix} \dot{m}_{ai}(k) - \dot{m}_{ao}(k) \\ \frac{1}{J}\left(\tau_{net}(k) - \tau_L(k)\right) \\ \frac{1}{\alpha_f}\left(\dot{m}_{fc}(k) - \dot{m}_f(k)\right) \\ \frac{1}{\alpha_e}\left(ST \times AFI(k) - T_{exh}(k)\right) \end{bmatrix} \tag{21}$$

where, $T$ is the sampling time. Figure 5 shows the schematic of the engine cold start second order DSMC with three control inputs and three outputs. The first controller is the $AFR$ controller which tracks the desired $AFR$ trajectory by adjusting the fuel injection rate. The second controller is the idle speed controller which tracks the desired engine speed profile via the air mass flow rate into the cylinders by using the air throttle body as the actuator. The third controller is $T_{exh}$ controller which tracks the desired exhaust gas temperature trajectory by regulating the spark timing. Figure 5 also shows that the feedback data from the engine are sampled and quantized via an analog-to-digital converter (ADC) unit. The control commands from the ECU are updated at a rate equal to $T$.

The analytic equations of the second order SISO DSMCs inputs can be obtained by applying Eq. (18) to the discretized nonlinear model in Eq. (21). However, it is important to incorporate any uncertainties in the engine model during the DSMC design. Otherwise, the controller can significantly deviate from its desired performance [22,23]. The discrete nonlinear equations of the engine model from Eq. (21) are updated to include unknown multiplicative uncertainty terms ($\varphi_i$) as follows:

$$\begin{bmatrix} m_a(k+1) \\ \omega_e(k+1) \\ \dot{m}_f(k+1) \\ T_{exh}(k+1) \end{bmatrix} = \begin{bmatrix} m_a(k) \\ \omega_e(k) \\ \dot{m}_f(k) \\ T_{exh}(k) \end{bmatrix} + T \underbrace{\begin{bmatrix} \varphi_{m_a}[-\dot{m}_{ao}(k)] \\ \varphi_{\omega_e}\left[\frac{1}{J}\left(-\tau_L(k)\right)\right] \\ \varphi_{\dot{m}_f}\left[\frac{1}{\alpha_f}\left(-\dot{m}_f(k)\right)\right] \\ \varphi_{T_{exh}}\left[\frac{1}{\alpha_e}\left(600 AFI(k) - T_{exh}(k)\right)\right] \end{bmatrix}}_{\varphi_i \times f_i(x(k))} + T \underbrace{\begin{bmatrix} \dot{m}_{ai}(k) \\ \frac{30000}{J} m_{a,d}(k) \\ \frac{1}{\alpha_f} \dot{m}_{fc}(k) \\ \frac{7.5}{\alpha_e}\Delta(k) \end{bmatrix}}_{g_i(x(k)) \times u_i(k)} \tag{22}$$

where $f_i(x(k))$ is the main part of the dynamics, which is subjected to uncertainties due to variation in the engine parameters or errors in estimating the model, and $g_i(x(k))$ is a non-zero input coefficient. By comparing Eq. (22) with Eq. (21), it can be easily concluded that the nominal values for $\varphi_i$ is "1". This means that if $\varphi_i \neq 1$, then the ultimate goal of the adaptation algorithm would be converging the unknown $\varphi_i$ to the nominal value, "1". We showed in [8] that solving the following adaptation law for each of the four controllers leads to $\varphi_i \rightarrow 1$:

$$\hat{\varphi}_i(k+1) = \hat{\varphi}_i(k) + \frac{T.s_i(k).f_i(x(k))}{\rho_{\varphi_i}} \tag{23}$$

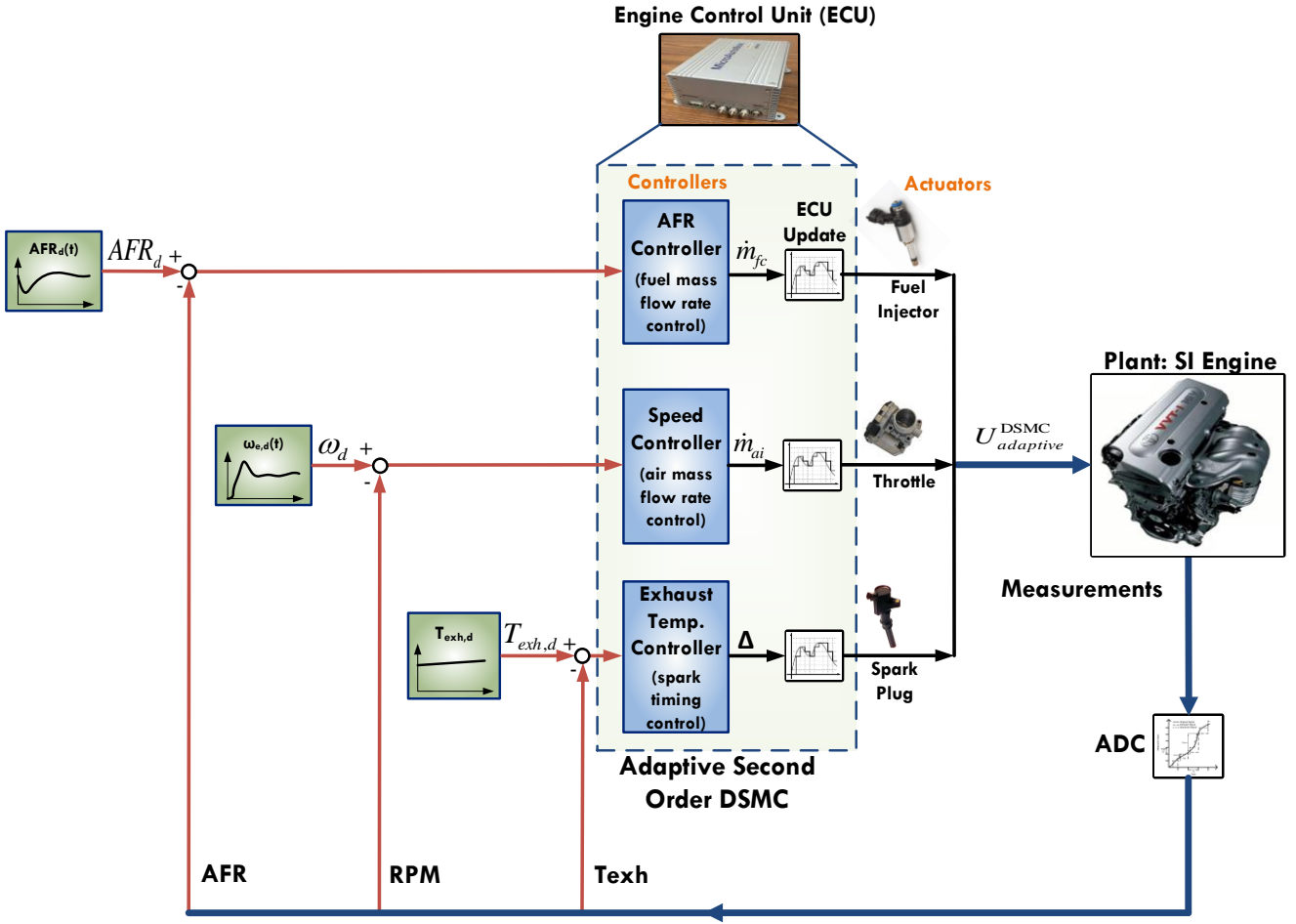

**Figure 5: Schematic of the second order adaptive DSMC for the engine cold start.**

where $\rho_{\varphi_i} > 0$ is a tunable adaptation gain chosen for the numerical sensitivity of the unknown multiplicative parameter estimation, and $\hat{\varphi}_i$ is the estimation of the unknown parameter $\varphi_i$. According to Eq. (22), four unknown uncertainty terms are considered in the modeled engine equations $(\varphi_{m_a}, \varphi_{\omega_e}, \varphi_{\dot{m}_f}, \varphi_{T_{exh}})$. $\varphi_{m_a}$ represents any error in the air mass dynamics, $\dot{m}_{ao}(k)$, which is calculated with respect to Eq. (A5) and is a direct function of the engine volumetric efficiency $(\eta_{vol})$. $\eta_{vol}$ curve is found by using Eq. (A6) in the Appendix. $\varphi_{\omega_e}$ compensates for any error in reading the torque map $(\tau_L)$ in the engine rotational dynamics. Fuel flow rate dynamics is a strong function of fuel evaporation time constant $(\alpha_f)$. $\varphi_{\dot{m}_f}$ represents any errors in estimating the fuel evaporation time constant. Finally, in the exhaust gas temperature dynamics, the exhaust gas time constant $(\alpha_e)$ plays an important role. Thus $\varphi_{T_{exh}}$ is representing any errors or variation in this time constant. Overall, the unknown multiplicative parameters inside the engine model can be estimated ($\hat{\varphi}_{m_a}, \hat{\varphi}_{\omega_e}, \hat{\varphi}_{\dot{m}_f}, \hat{\varphi}_{T_{exh}}$) by solving Eq. (23) for each controller, simultaneously. Finally, the control inputs of the SISO adaptive second order DSMC in the presence of modeling uncertainties become:

$$\dot{m}_{ai}(k) = \frac{1}{T}[\hat{\varphi}_{m_a}(k)\dot{m}_{ao}(k)T - (\beta_4(k)+1)s_4(k) + m_{a,d}(k+1) - m_{a,d}(k)] \tag{24}$$

$$m_{a,d}(k) = \frac{J}{30000T}[\hat{\varphi}_{\omega_e}(k)\frac{T}{J}(100 + 0.4\omega_e(k)) - (\beta_2(k)+1)s_2(k) + \omega_{e,d}(k+1) - \omega_{e,d}(k)] \tag{25}$$

$$\dot{m}_{fc}(k) = \frac{\alpha_f}{T}\left[\hat{\varphi}_{\dot{m}_f}(k)\frac{T}{\alpha_f}\dot{m}_f(k) - (\beta_1+1)s_1(k) + \dot{m}_{f,d}(k+1) - \dot{m}_{f,d}(k)\right] \tag{26}$$

$$\Delta(k) = \frac{\alpha_e}{7.5.AFI.T}[-\hat{\varphi}_{T_{exh}}(k)\frac{T}{\alpha_e}(600.AFI - T_{exh}(k)) - (\beta_3+1)s_3(k) + T_{exh,d}(k+1) - T_{exh,d}(k) \tag{27}$$

In the absence of modeling uncertainties ($\varphi_i = 1$), Figure 6-a shows the SISO second order DSMC performance in tracking the variable and non-smooth *AFR*, exhaust gas temperature, and engine speed trajectories under sampling time of 20 *ms* and quantization level of 16 *bit*. According to the RGA analysis results in Figure 4, the highest closed-loop coupling, outside the main closed-loop, exists between *AFR* and engine speed control loops. As can be seen from Figure 6-a, whenever the engine speed decreases sharply, the *AFR* tracking performance is affected. However, the SISO controller handles this coupling, which acts as a disturbance on the *AFR* controller, accordingly, and the disturbance effects on the *AFR* controller performance is less than 0.5%, in terms of the tracking errors. Overall, Figure 6 illustrates that for the engine cold start problem, despite the non-smooth shape of the desired trajectories, the SISO second order DSMCs are able to track all the desired trajectories accurately, without getting affected by the other control loops considerably. Thus, the SISO controller design can fulfill the desired performance requirements of the complex engine cold start dynamics.

Figure 7 shows the performance of the adaptation laws for removing up to 50% uncertainty within the engine dynamics. This signifies the importance of handling the unknown uncertainty terms. As can be seen from Figure 7-(a,b,c,d)$_1$, in the absence of the adaptation, the multiplicative unknown terms generate a large offset in the engine dynamics ($f_i(x(k))$). These errors within the model will propagate through the controller structure and lead to controller failure. However, when the adaptation becomes active, the adapted models converge to the nominal models once the adaptation period is over. Figure 7-(a,b,c,d)$_2$ show that the adaptation mechanism is able to converge all the four unknown multiplicative parameters to their nominal values ("1") in less than 5 *sec* of the simulation time. Overall, the adaptation mechanism is able to remove up to 95% of the uncertainties in the models in comparison with the non-adaptive models.

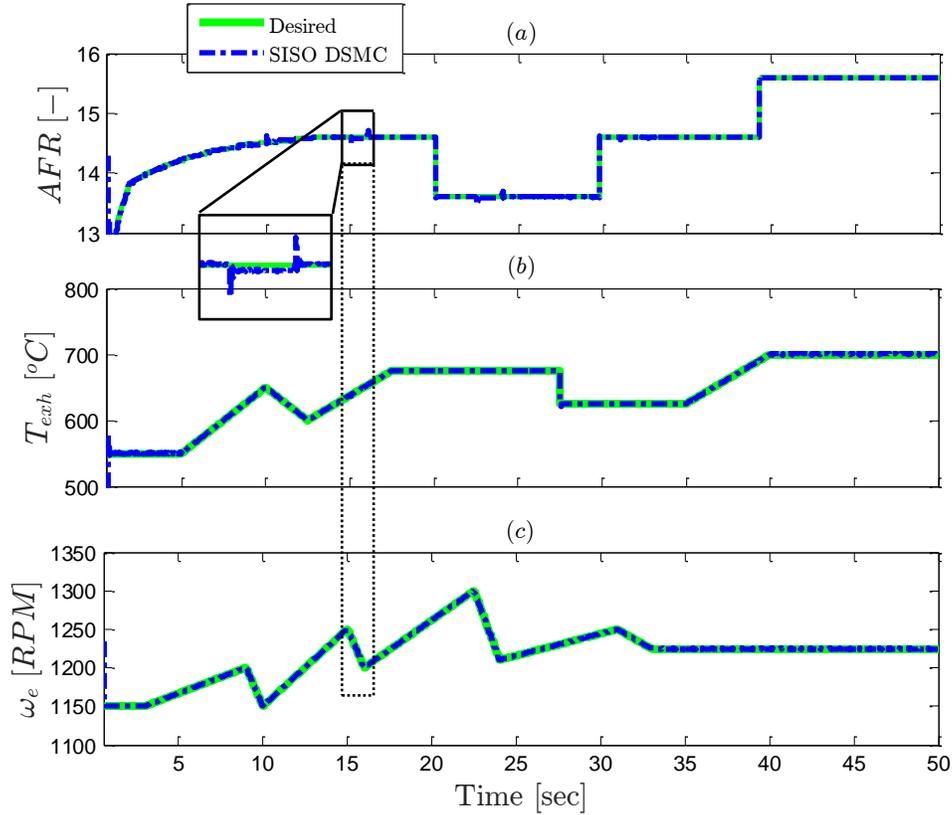

**Figure 6: Results of tracking desired engine trajectories by using the designed SISO DSMC**
**(*T*=20 ms, quantization level=16 *bit*).**

## 5- Controller Real-Time Verification

In this section, the designed SISO adaptive second order DSMC for the engine cold start problem is tested in real-time by using a processor-in-the-loop (PIL) setup with an actual ECU. The schematic of the PIL setup is shown in Figure 8. The PIL setup has two different processors which communicate with each other. The first processor is a National Instrument (NI) PXI processor (NI PXIe-8135). The PXI processor is used to implement the model of the engine cold start emission, via NI VeriStand® software, and sends feedback signals from the model to the ECU. The second processor is a dSPACE MicroAutoboxII (MABX), which is considered as the ECU. The feedback signals from the engine model (PXI processor) and the control commands at the ECU outputs are sampled every 20 *ms*, and digitized with a quantization level of 16 *bit*. The second order DSMCs along with the adaptation mechanism are implemented into the MABX via dSPACE Control Desk®. The real-time tracking performance of the second order DSMCs and the engine and catalytic converter performances are monitored via a user interface desktop computer.

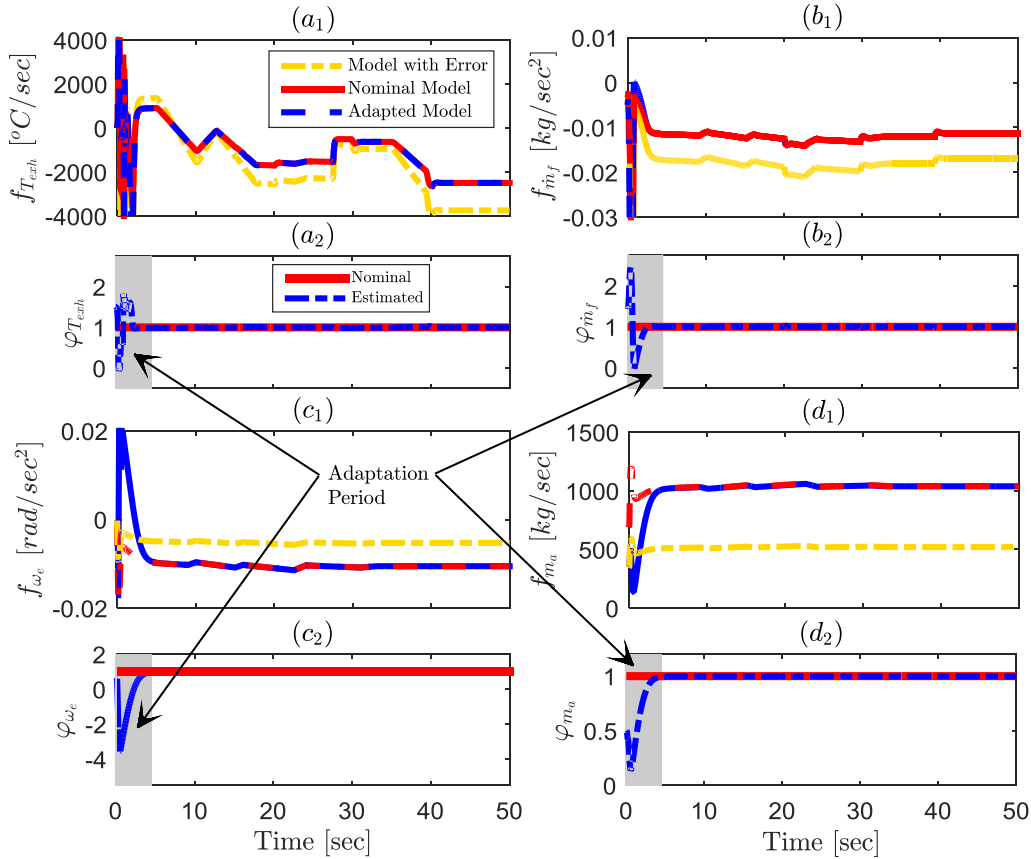

**Figure 7: The impacts of multiplicative uncertainty terms on the engine dynamics inside the second order SISO DSMC and how the adaptation mechanism drives the model with error to its nominal value: (a) $T_{exh}$, (b) $\dot{m}_f$ , (c) $\omega_e$, and (d) $m_a$ ($T$=20 ms, quantization level=16-$bit$).**

Figure 9 shows the tracking performance of the SISO adaptive and non-adaptive second order DSMCs under modeling uncertainties. In the absence of the adaptation, it can be seen that the controller fails to track the desired trajectories. The reason for this failure is the errors within the engine model which were shown in Figure 7. On the other side, once the adaptation period (the first 5 *sec*) is over, the errors in the model are completely removed; then, the adaptive DSMC tracks the desired trajectories accurately and smoothly. The real-time test results in Figure 9 show that the adaptive controller can reduce the tracking errors by more than 90%, in terms of the mean tracking errors, in comparison with the non-adaptive DSMC.

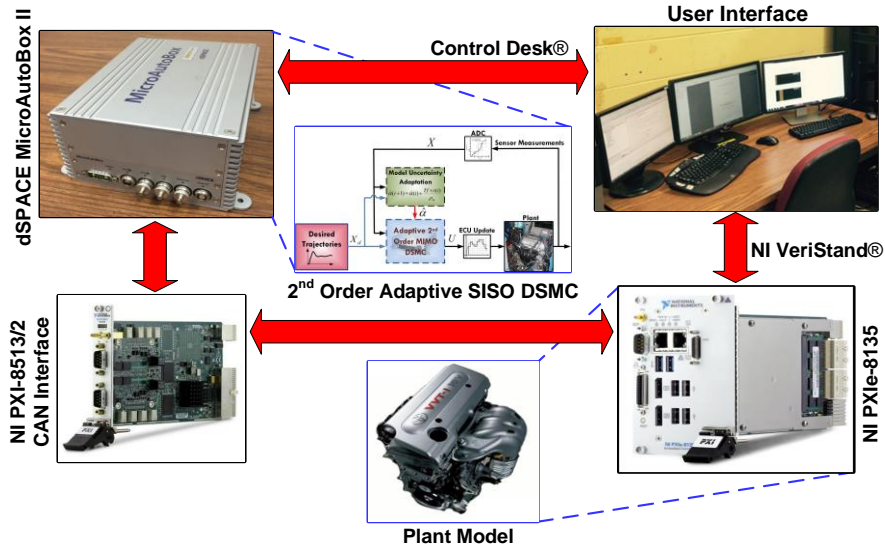

**Figure 8: Schematic of the processor-in-the-loop (PIL) setup for real-time verification of the designed SISO DSMCs.**

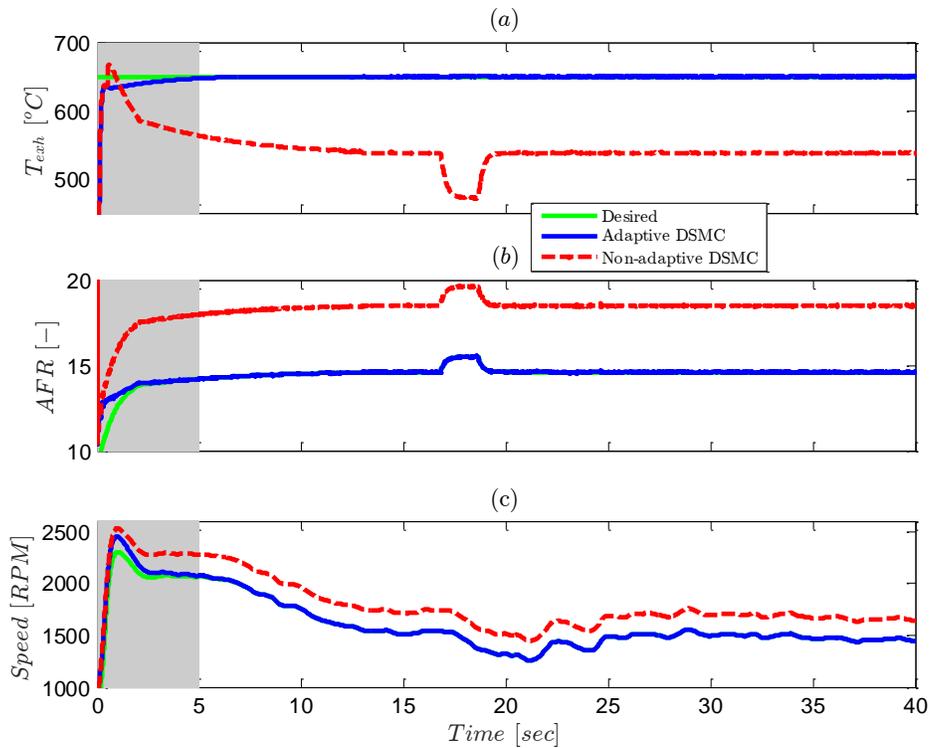

**Figure 9: Results of engine control under model uncertainties: (a) exhaust gas temperature, (b) *AFR*, and (c) engine speed (*T*=20 ms, quantization level=16 *bit*).**

The desired trajectories in Figure 9 are taken from Toyota ECU, and used as the optimum trade-off among engine speed, *AFR*, and exhaust gas temperature during the engine cold start phase. This means that accurate tracking of these trajectories by the adaptive DSMC leads to fast catalyst light-off, and low tailpipe HC emissions. Figure 10 illustrates the real-time performance of the control system in minimizing the tailpipe HC emission. The tailpipe and engine-out HC flow rates are both shown in Figure 10-a. As can be seen, during the first seconds of the cold start period, both engine-out and tailpipe HC emissions are high. This is due to low catalyst temperature (Figure 10-c) and rich air-fuel mixture used to avoid combustion instability at the beginning of the cold start period. After *AFR* becomes stoichiometric and the catalytic converter reaches the light-off temperature, the tailpipe HC emission rate starts to fall. Eventually, at the end of the test period (i.e., 40 *sec*), the tailpipe HC emission rate approaches to zero.

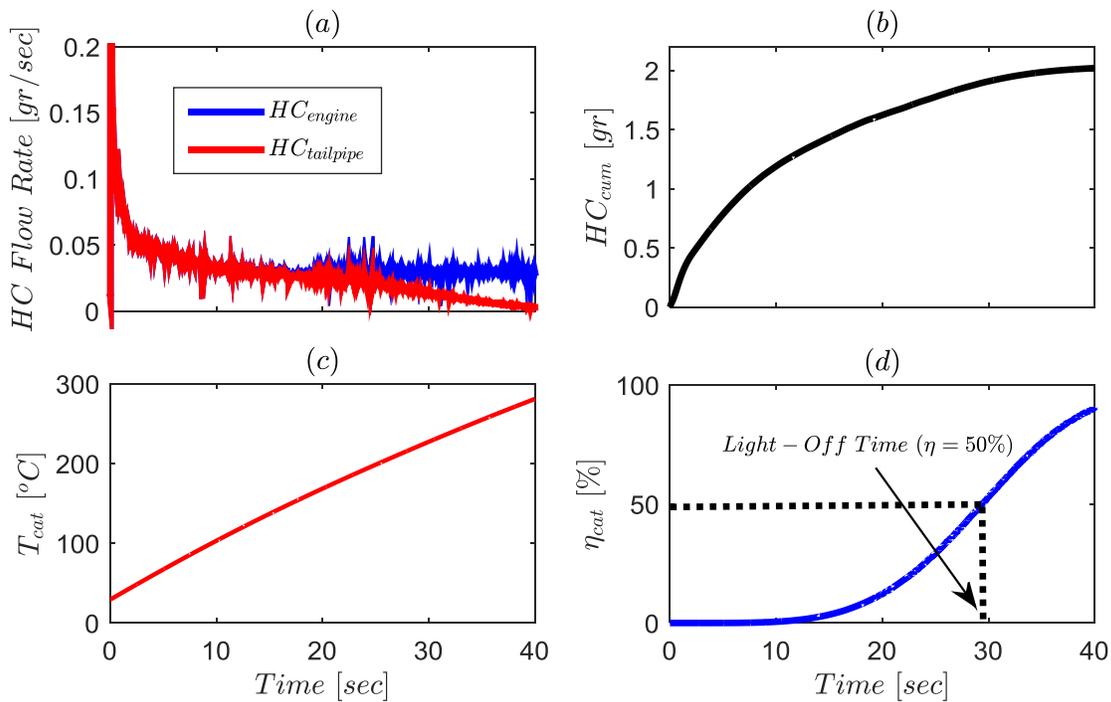

**Figure 10: Cold start emission control results by using SISO second order adaptive DSMC: (a) engine-out and tailpipe HC emission flow rates, (b) cumulative tailpipe HC emission, (c) catalyst temperature, and (d) catalytic converter efficiency (*T*=20 ms, quantization level=16-*bit*).**

Figure 10-b shows the cumulative amount of the tailpipe HC emission generated during the cold start phase. The limit for the cumulative HC emission for North America UDDS driving cycle at the end of the first 40 *sec* of the cold start phase is 2.5 *gr*, assuming that over 80% of the total HC emissions come from the cold start period [2,3]. As can be seen from Figure 10-b, under the data sampling and quantization imprecisions, and also uncertainties within the engine model, the second order adaptive SISO DSMC is able to meet the HC emission limit (2.5 *gr*) by end of the test period (~2 *gr*). Figure 10-

c shows that the catalytic converter reaches to 50% efficiency (light-off) in 30 *sec*. In a similar manner, Figure 10-d shows that without using any external heating energy sources, catalyst conversion efficiency of more than 90% can be achieved by using the adaptive second order DSMC.

## Summary and Conclusion

In this paper, a novel second order DSMC with adaptation against modelling uncertainties, and consideration of the data sampling and quantization imprecisions, was designed for an SI engine cold start HC emission problem. The structure of the closed-loop controller was chosen based on a dynamic coupling analysis by using relative gain array (RGA) analysis. The RGA analysis showed that despite the complex and internally coupled dynamics of the engine during the cold start phase, the second order DSMC can be designed in a decentralized (SISO) structure with separate control loops for *AFR*, engine speed and exhaust gas temperature. Simulation and real-time experimental verification results of the designed controller showed that:

- A set of three SISO DSMCs can track desired *AFR*, engine speed and exhaust gas temperature trajectories accurately during cold start phase.

- In the presence of up to 50% modelling uncertainty within each of the SISO controllers, the adaptation mechanism inside the second order DSMC can remove the errors in the model permanently in less than 5 *sec* by up to 95%. Moreover, the comparison results between adaptive and non-adaptive DSMCs showed that the adaptive DSMC can improve the tracking performance of the non-adaptive DSMC by up to 90%.

- The adaptive DSMC is able to heat-up the catalyst to reach to 50% efficiency in less than 30 *sec*. This is the outcome of accurate tracking of optimum trajectories that lead to shorten catalyst light-off time and reaching over 90% conversion efficiency for the catalytic converter at the end of test period (40 *sec*).

## Acknowledgment

This material is based upon the work supported by the US National Science Foundation (NSF) under Grant No. 1434273. Prof. Karl Hedrick from University of California-Berkeley, and Dr. Ken Butts from Toyota Motor Engineering & Manufacturing North America are gratefully acknowledged for their technical feedback during the course of this study.

## Appendix

*Engine cold start model*:

**Table A1. Plant Model Constants**

| Constant | Value [unit] |
|---|---|
| $J$ | $0.1454 \, [m^2 kg]$ |
| $\alpha_f$ | $0.06 \, [1/sec]$ |
| $mC_p$ | $1250 \, [J/K]$ |
| $a$ | $-2 \, [-]$ |
| $n$ | $5 \, [-]$ |
| $\theta_{evo}$ | $110 \, [^\circ \text{ATDC}]$ |
| $r_c$ | $9 \, [-]$ |

*Plant Model Functions:*

$$ST = 7.5\Delta + 600 \tag{A1}$$

$$AFI = \cos\{0.13[AFR(m_a, \omega_e, \dot{m}_f) - 13.5]\} \tag{A2}$$

$$\tau_{net}(\omega_e, m_a) = 30000m_a - 0.4\omega_e - 100 \tag{A3}$$

$$\alpha_e = \frac{2\pi}{\omega_e} \tag{A4}$$

$$\dot{m}_{ao} = 0.0254\eta_{vol}m_a\omega_e \tag{A5}$$

$$\begin{aligned} \eta_{vol} = \; & m_a^2(-0.1636\omega_e^2 - 7.093\omega_e - 1750) \\ & + m_a(0.0029\omega_e^2 - 0.4033\omega_e + 85.38) \\ & - (1.06 \times 10^{-6}\omega_e^2 - 0.0021\omega_e - 0.2719) \end{aligned} \tag{A6}$$

$$\dot{Q}_{in} = 16(T_{exh} - T_{cat}) \tag{A7}$$

$$\dot{Q}_{out} = 0.642(T_{cat} - T_{atm}) \tag{A8}$$

$$\dot{Q}_{gen} = 22.53(\dot{m}_{ao} + \dot{m}_f T_{exh})\eta_{cat}\dot{H}C_{eng} \tag{A9}$$

$$\theta_0 = \Delta + 10 \tag{A10}$$

$$\begin{aligned} & \delta\theta = k_1(AFR - 16.2)^2 + k_2, \\ where \; & \begin{cases} K_1 = 0.1; K_2 = 80 \; AFR > AFR_{st} \\ K_1 = 0.4; K_2 = 80 \; AFR < AFR_{st} \end{cases} \end{aligned} \tag{A11}$$